\documentclass[aps,pra,preprint]{revtex4}
\usepackage{graphicx}
\usepackage{dcolumn}
\usepackage{amsmath}
\usepackage{amssymb}
\usepackage{ulem}
\usepackage{color}

\def\bef{\begin{framed}}
\def\eef{\end{framed}}
\def\be{\begin{equation}}
\def\ee{\end{equation}}
\def\ber{\begin{eqnarray}}
\def\eer{\end{eqnarray}}

\def\rv{{\bf r}}

\def\jv{{\bf j}}

\def\Av{{\bf A}}

\begin{document}
\title{Comment on\\
 ``Density and Physical Current Density Functional Theory"\\
by Xiao-Yin Pan and Viraht Sahni\\ Int. J.   Quant. Chem.  {\bf 110}, 2833 (2010)}
\author{G. Vignale and C. A. Ullrich}
\affiliation{Department of Physics, University of Missouri-Columbia,
Columbia, Missouri 65211}
\author{K. Capelle}
\affiliation
{Centro de Ci\^encias Naturais e Humanas, Universidade Federal do ABC (UFABC), Santo Andr\'e 09210-170, SP, Brazil}

\date{\today}
\maketitle
Among the extensions of density functional theory (DFT), the non-relativistic current density functional theory (CDFT), designed by Vignale and Rasolt (VR)~\cite{VR,VR2} for electronic systems in a static magnetic field, stands out for its conceptual subtlety.
Electrons couple to a magnetic vector potential $\Av(\rv)$ via the so-called {\it paramagnetic} current density $\jv_p(\rv)$, which  in the non-relativistic theory  differs from the {\it physical} current density $\jv (\rv)$ and is related to the latter by
\be\label{Currents}
\jv(\rv)=\jv_p(\rv)+\rho(\rv) \Av(\rv)
\ee
[Here, $\rho(\rv)$ is the ground-state density, and we use units in which $e=m=c=1$].
It has been proved~\cite{VR} that $\rho(\rv)$ and $\jv_p(\rv)$ uniquely determine (through a one-to-one correspondence) the ground-state wave function, but not necessarily the external scalar and vector potentials, $V(\rv)$ and $\Av(\rv)$,  that give rise to them.  Schematically,  one has
\be\label{VR}
{\rm VR:}~~~~~~~\{\rho,\jv_p\} \leftrightarrow \psi \leftarrow \{V,\Av\}\,,
\ee
where the last arrow points only one way.
Although the external potentials are not uniquely determined by the densities -- {a fact that was overlooked in Ref.~\onlinecite{VR} and} was first pointed out by Capelle and Vignale in Ref.~\onlinecite{CV} --  the one-to-one correspondence between densities and ground-state wave functions is sufficient to construct  the universal ``internal energy" functional $F[\rho,\jv_p]$ and the exchange-correlation energy functional $E_{xc}[\rho,\jv_p]$, and the constrained search algorithm and the variational principle remain unaffected~\cite{KSU}.

In spite of this,  a recent paper by Pan and Sahni (PS)~\cite{PS}   seizes on the non-uniqueness of the potentials -- {the point of  Ref.~\onlinecite{CV}} --  as a {\it casus belli} against the VR version of CDFT.  The authors of this paper have stated that (i) the ground-state wave function ``cannot be a functional of $\rho$ and $\jv_p$" (end of Section 6) and (ii)  the physical current $\jv$ is the ``natural basic variable" for CDFT \cite{Diener}.

Statement (i) is obviously a mistake: the ground-state {\it wave function} is indeed a unique functional of $\rho$ and $\jv_p$ [see Eq.~(\ref {VR}) above], {as demonstrated in Refs.~\onlinecite{VR} and \onlinecite{CV}}.  In this comment we show that statement (ii) is also erroneous. The physical-current density version of CDFT leads to problems far more serious than the ones it purports to cure.  Let us see why.

The essential difficulty with the choice of the physical current $\bf j$ as a basic variable is that the correspondence between density and wave functions, which is one-to-one in the VR formulation,  becomes many-to-one in the PS formulation.   Schematically, {PS have proved}
\be\label{PS}
{\rm PS:}~~~~~~~\{\rho,\jv\} \leftrightarrow  \{V,\Av\} \rightarrow \psi \,,
\ee
where the last arrow points only one way.  {We believe this statement to be correct, since it follows immediately from previously proved results~\cite{Diener}}. 
{However, different current densities may now be associated with one and the same ground-state wave function.}  Indeed, given a wave function, it is impossible to work out the physical current without the additional knowledge of the vector potential.   The consequences of this fact are disturbing,  as we shall now show.

Let us follow PS and write the energy functional as
\be\label{PSFunctional}
E_{V,\Av}[\rho',\jv']=F_{PS}[\rho',\jv']+\int \jv'(\rv)\cdot \Av(\rv) d\rv -\frac{1}{2} \int \rho'(\rv) A^2(\rv) d\rv +\int\rho'(\rv)V(\rv) d\rv \,,
\ee
where $F_{PS}[\rho,\jv]$ is the expectation value of the standard many-body Hamiltonian $\hat T+\hat U= -\frac{1}{2}\sum_i \nabla_i^2 + \sum_{i<j} U(\rv_i-\rv_j)$ in the ground-state wave function, which is uniquely determined by $n(\rv)$ and $\jv(\rv)$ according to~(\ref{PS}).  We note that, contrary to what is  stated in Eq. (59) of PS, the functional $F_{PS}[\rho',\jv']$ {\it cannot} be constructed from a constrained minimization of $\hat T +\hat U$ on the set of wave functions that yield $\rho$ and $\jv$.  The reason is that  this set is undefined, unless the vector potential is  specified -- and therefore Eq. (59) of Ref.~\onlinecite{PS} is meaningless. 

The standard variational principle for ground-state wave functions mandates that we determine the ground-state values of $\rho$ and $\jv$ by minimizing $E_{V,\Av}[\rho',\jv']$ with respect to $\rho'$ and $\jv'$ {\it at constant $V$ and $\Av$}.
Notice, however, that $\jv=\jv_p+\rho\Av$ [Eq.~(\ref{Currents})].    From the form of this relation we clearly see that {\it a minimization with respect to $\jv$ and $\rho$ at constant $\Av$  and $V$ is equivalent to a minimization with respect to $\jv_p$ and $\rho$ under the same conditions}.  Once $\Av$ is fixed, $\jv$ and $\jv_p$ become essential the same variable, and we may as well rewrite the functional to be minimized in terms of $\jv_p$ as follows:
\be\label{VRFunctional}
E_{V,\Av}[\rho',\jv_p']=F_{VR}[\rho',\jv_p']+\int \jv_p'(\rv)\cdot \Av(\rv) d\rv +\frac{1}{2} \int \rho'(\rv) A^2(\rv) d\rv +\int\rho'(\rv)V(\rv) d\rv
\ee
(notice the change in sign of the term containing $A^2$).  $F_{VR}[\rho',\jv_p']$ is precisely the universal VR energy functional of $\jv_p$.  Thus, the minimization of the PS functional with respect to $\jv$ at constant $\Av$ is equivalent to the minimization of the VR functional with respect to $\jv_p$ at constant $\Av$, if one takes proper care of the $\Av$ that is hidden in the definition of $\jv$.  Under the stipulation of constant $\Av$, the PS theory is nothing more than a garbled reformulation of the VR theory.

However, in the above discussion we took a lenient view of the functional~(\ref{PSFunctional}).  A  stricter interpretation,
as advocated by PS, would be to treat $\jv$ as the ``natural basic variable" and perform the minimization with respect to it, only keeping constant  the $\Av$ that {\it explicitly} appears in Eq.~(\ref{PSFunctional}).
By doing this, we run into a serious problem.  We know from Ref.~\onlinecite{CV} that it is possible to find two different sets of infinitesimally close densities, $\{\rho(\rv), \jv(\rv)\}$ and $\{\rho(\rv),\jv'(\rv)\equiv \jv(\rv)+\delta\jv(\rv)\}$, that correspond to the same ground-state wave function, and, therefore, yield the same $F$:
$F_{PS}[\rho,\jv]=F_{PS}[\rho,\jv']$.  Then in the infinitesimal variation from  $\{\rho(\rv), \jv(\rv)\}$ to $\{\rho(\rv), \jv'(\rv)\}$ {at constant $V$ and $\Av$ we have $\delta F_{PS}=0$ and $\delta \rho=0$ and} the full energy functional changes by
\be
\delta E_{V,\Av}=\int \delta \jv(\rv)\cdot \Av(\rv) d\rv \neq 0\,.
\ee
The energy functional is no longer stationary for infinitesimal variations about the chosen ground-state densities $\rho$ and $\jv$!
This completely spurious result has its root in the ignorance of the vector potential-dependence of $\jv$ {and indeed disappears as soon as one rewrites the energy functional in the equivalent VR form of Eq.~(\ref{VRFunctional}), for which  $\delta F_{VR}$ and $\delta \jv_p$ are both zero}.  {But, the paradox shows} how dangerous it is to work with a basic variable that is not an intrinsic property of the system, and therefore cannot be uniquely determined by the wave function alone.

To summarize, we have found that the PS formulation of CDFT is, at  best, a garbled reformulation of the VR theory, and, at  worst,  a potential source of mistakes insofar as it complicates the formulation of the variational principle and prevents the constrained search construction of the universal functional.


\begin{thebibliography}{100}
\bibitem{Diener} G. Diener, in J. Phys: Condensed Matter {\bf 3}, 9417 (1991). In this paper Diener used the classic Hohenberg-Kohn contradiction to prove that if two different sets of potentials $\{V,\Av\}$ and $\{V',\Av'\}$ have different ground-states, then the corresponding sets of densities $\{\rho,\jv\}$ and $\{\rho',\jv'\}$ cannot coincide.  Missing from his analysis (as also from the VR analysis of Ref.~\onlinecite{VR}) was the possibility that  $\{V,\Av\}$ and $\{V',\Av'\}$ might have the same ground-state.  But, in this case, we have $\jv_p=\jv_p'$  and $\rho=\rho'$,  and since the vector potentials must necessarily be different~\cite{CV} we immediately conclude that $\jv \neq \jv'$.
\end{thebibliography}
\end{document}